\documentclass[a4paper,12pt]{article}
\usepackage[utf8]{inputenc}
\usepackage{amssymb}
\usepackage{graphicx,amssymb,amsfonts}
\newlength{\extraspace}
\newlength{\extraspaces}
\RequirePackage[T1]{fontenc}
\RequirePackage{color,soul}
\RequirePackage{amssymb}
\RequirePackage[figtopcap]{subfigure}
\RequirePackage{array,multirow,graphicx}
\newcommand{\be}{\begin{equation}
\addtolength{\abovedisplayskip}{\extraspaces}
\addtolength{\belowdisplayskip}{\extraspaces}
\addtolength{\abovedisplayshortskip}{\extraspace}
\addtolength{\belowdisplayshortskip}{\extraspace}}
\newcommand{\ee}{\end{equation}}

\newcommand{\ba}{\begin{eqnarray}
\addtolength{\abovedisplayskip}{\extraspaces}
\addtolength{\belowdisplayskip}{\extraspaces}
\addtolength{\abovedisplayshortskip}{\extraspace}
\addtolength{\belowdisplayshortskip}{\extraspace}}
\newcommand{\ea}{\end{eqnarray}}

\newcommand{\nonu}{\nonumber \\[.5mm]}
\newcommand{\A}{&\!\!\!}

\newcommand{\newsection}[1]{
\vspace{7mm} \pagebreak[3] \addtocounter{section}{1}
\setcounter{subsection}{0} \setcounter{footnote}{0}
\begin{center}
{\large {\bf \thesection. #1}}
\end{center}
\nopagebreak
\medskip
\nopagebreak \hspace{3mm}}

\begin{document}
\begin{center}
{ \it \bf \large Generalized teleparallel cosmology\\
and  \\ initial singularity crossing
}\\
\bigskip
\bigskip
\bigskip
{  Adel Awad$^{1,2}$  and Gamal Nashed$^{1,3}$ }

{ \it \small $^1$Center for Theoretical Physics, British University of Egypt,
Sherouk City 11837, Egypt\\
$^2$Department of Physics, Faculty of Science, Ain Shams University, Cairo 11566, Egypt\\
$^3$Department of Mathematics, Faculty of Science, Ain Shams University, Cairo 11566, Egypt\\}
\end{center}
\bigskip

\date{}

We present a class of cosmological solutions for a generalized teleparallel gravity with $f(T)=T+\tilde{\alpha} \, (-T)^n$, where $\tilde{\alpha} $ is some parameter and $n$ is an integer or half-integer. Choosing $\tilde{\alpha} \sim G^{n-1}$, where $G$ is the gravitational constant, and working with an equation of state $p=w\, \rho$, one obtains a cosmological solution with multiple branches. The dynamics of the solution describes standard cosmology at late times, but the higher-torsion correction changes the nature of the initial singularity from big bang to a sudden singularity. The milder behavior of the sudden singularity enables us to extend timelike or lightlike curves, through joining two disconnected branches of solution at the singularity, leaving the singularity traversable. We show that this extension is consistent with the field equations through checking the known junction conditions for generalized teleparallel gravity. This suggests that these solutions describe a contracting phase a prior to the expanding phase of the universe.

\begin{center}
\newsection{\bf Introduction}
\end{center}

The geometric construction of the teleparallel equivalent of general relativity (TEGR) \cite{AP, APVB, Mj} is given by a Weitzenb\"ock spacetime. This spacetime has a nonsymmetric connection which yields a non vanishing torsion, but a vanishing Riemann curvature. Alternatively, the spacetime of Einstein general relativity (GR) has a symmetric Levi-Civita connection that is torsionless and yields a Riemannian curvature. The torsion scalar $T$ in TEGR plays an essential role which is very similar to the role of curvature in GR. Although both theories have a different geometrical structures, they are equivalent theories regarding both the field equations and equation of motion of test particles.

In recent years there has been lots of interest in generalizing TEGR \cite{FF}--\cite{BS}. Many features of generalized teleparallel gravity have been considered in literature especially that describing cosmology with dark energy/matter through torsion. Furthermore, several interesting black holes solutions have been found and studied, for example see Ref's. \cite{CC}--\cite{CC23}. It is important to stress that $f(T)$ has extra degrees of freedom \cite{LSB} regarding TEGR (and, thus, to GR). These extra degrees of freedom are linked to the fact that the field equations are not invariant under local Lorentz transformations \cite{LSB, LSB1}. Consequently, there exists a special global reference frame defined by the autoparallel curves of the manifold that solve the field equations. For example, as investigated in \cite{TC}, a diagonal tetrad is not the best choice for non-flat homogenous and isotropic cosmological solutions (Friedman-Lemaitre-Robertson-Walker universes) and for spherically symmetric solutions (Schwarzschild or Schwarzschild-de Sitter solutions).

The reconstruction of a suitable $f(T)$ gravitational theories, which describe the early Universe
inflation as well as late-time acceleration, was studied in detail in  \cite{CC4,WNA}--\cite{WN1}.   A reconstructed  different forms $f(T)$ have been obtained using entropy-corrected versions of the holographic \cite{PD16}.

One of the interesting models of $f(T)$ theories is the choice $f(T)=T+\tilde{\alpha} \,(- T)^n$. Spherically symmetric solutions for these models with $n=2$ in the weak-field approximation, has been studied in details \cite{GG}--\cite{XD}. These solutions are interesting and their higher-curvature corrections can be constrained using solar system data. In a recent paper \cite{RR} a spherically symmetric solutions were studied, in the weak-field approximation, for Lagrangians in the form $f(T) = T + \tilde{\alpha} (-T)^n$, with $ n \neq 1$, and solved the field equations using a non diagonal tetrad.  It is shown that, to the lowest order, the perturbations of the corresponding GR solutions are in the form of power laws $\propto \tilde{\alpha}  r^{2-2n}$. More recently,  cosmic chronometers data and observations related to the variation of fundamental constants are used to impose constraints on the viable $f(T)$ gravity models \cite{NPS, NBPS}. Also, a gravitational baryogenesis scenario in a universe governed by $f(T)$ gravitational theories was studied in \cite{OS}.

The aim of  this article is to present a class of new cosmological solutions for $f(T)$ theory with $f(T)=T+\tilde{\alpha} \,(- T)^n$, using the usual equation of state $p=w\, \rho$, where $\tilde{\alpha}$ is some parameter and $n$ is an integer or half integer. These solutions approach the standard cosmology at late times. Our analysis show that the higher-curvature corrections to the TEGR theory changes the nature of the initial singularity from big bang to a sudden singularity which leads to a less singular behavior and enable us to extend nonspacelike curves beyond the singular point. A sudden singularity is characterized by a finite scale factor $a$ and $\dot{a}$, but a divergent $\ddot{a}$ \cite{barrow-s,barrow-s1,barrow-s2}.  Cosmologists are interested in sudden singularities since they appear naturally in several models of dark energy and unified dark fluid (for examples see \cite{cosmod1}--\cite{cosmod3}). Sudden singularities are among possible future-time singularities for FRW cosmologies that have been analyzed in \cite{sing-class}. One of the interesting features of spacetimes with a sudden singularity in GR is that its geodesics can be extended across the singular region \cite{jambrina}--\cite{barrow-g}, therefore, this singularity is traversable.  It is known that geodesic equations in GR are completely equivalent to equations of motion of test particles in teleparallel gravity, therefore, it is natural to define a spacetime singularity in teleparallel gravity as points where timelike or lightlike curves of test particles motion end. This definition is equivalent to geodesic incompleteness in GR. As a consequence of that nonspacelike curves of test particles can not be extended across big bang singularity of teleparallel gravity with $f(T)=T$, which is consistent with the equivalence between GR and TEGR.  Keeping in mind this defining feature of a singularity, we show that curves of test particles can be extended across the singular point leaving these new solutions traversable. Furthermore, we show that these new extensions are consistent with the field equations through checking the junction conditions of generalized teleparallel gravity that have been derived in Ref. \cite{DDG}.

The rest of the paper is organized as follows; in section 2 we give a brief review of $f(T)$ theories. In section 3, we present the exact cosmological solution for general $n$, then gives some details about two representative cases; $n=3/2$ and $n=2$ analyzing possible branches of solutions. In section 4, we derive the general solution near a singularity (for integer "n"), extend the spacetime beyond the singular point, then check the consistency of this spacetime extension with known junction condition derived for $f(T)$ in the literature. In section 5, we show how timelike and lightlike curves of test particles can be extended beyond the singular point leaving the singularity traversable. In section 6, we give our conclusion.

\newsection{Brief review of f(T)}
In the Weitzenb\"ock space time, the fundamental field
variables describing gravity are a quadruplet of parallel vector
fields \cite{Wr} ${e_i}^\mu$, which we call the tetrad field. This is
characterized by the auto parallel condition:  \begin{equation} D_{\nu} {e_i}^\mu=\partial_{\nu}
{e_i}^\mu+{\Gamma^\mu}_{\lambda \nu} {e_i}^\lambda=0, \end{equation} where
${\Gamma^\mu}_{\lambda \nu}$ defines the nonsymmetric affine
connection:  \begin{equation} {\Gamma^\lambda}_{\mu \nu} \stackrel{\rm def.}{=}
{e_i}^\lambda {e^i}_{\mu, \nu}, \end{equation} with $e_{i \mu, \
\nu}=\partial_\nu e_{i \mu}$.

Equation (1) leads to the metricity  condition and the identical
vanishment  of the curvature tensor defined by
${\Gamma^\lambda}_{\mu \nu}$, given by equation (2). The metric
tensor $g_{\mu \nu}$
 is defined by
\begin{equation} g_{\mu \nu} \stackrel{\rm def.}{=}  \eta_{i j} {e^i}_\mu {e^j}_\nu, \end{equation}
with $\eta_{i j}=(-1,+1,+1,+1)$ being the metric of Minkowski
space time. We note that, associated with any tetrad field
${e_i}^\mu$ there  is a metric field defined
 uniquely by (3), while a given metric $g^{\mu \nu}$ does not
 determine the tetrad field completely;  any local Lorentz
 transformation of the tetrads ${e_i}^\mu$ leads to a new set of
 tetrads which also satisfy (3).

The torsion components and the contortion are defined as: \begin{eqnarray}
{T^\alpha}_{\mu \nu}  &\stackrel {\rm def.}{=} &
{\Gamma^\alpha}_{\nu \mu}-{\Gamma^\alpha}_{\mu \nu} ={e_a}^\alpha
\left(\partial_\mu{e^a}_\nu-\partial_\nu{e^a}_\mu\right),\nonumber\\
{K^{\mu \nu}}_\alpha  & \stackrel {\rm def.}{=} &
-\frac{1}{2}\left({T^{\mu \nu}}_\alpha-{T^{\nu
\mu}}_\alpha-{T_\alpha}^{\mu \nu}\right), \end{eqnarray}   where the contortion
equals the difference between Weitzenb\"ock  and Levi-Civita
connection, i.e., ${K^{\mu}}_{\nu \rho}= {\Gamma^\mu}_{\nu \rho
}-\left \{_{\nu  \rho}^\mu\right\}$.

 The skew symmetric tensor ${S_\alpha}^{\mu \nu}$ is defined as: \begin{equation} {S_\alpha}^{\mu \nu}
\stackrel {\rm def.}{=} \frac{1}{2}\left({K^{\mu
\nu}}_\alpha+\delta^\mu_\alpha{T^{\beta
\nu}}_\beta-\delta^\nu_\alpha{T^{\beta \mu}}_\beta\right). \end{equation}  The torsion scalar is
defined as \begin{equation} T \stackrel {\rm def.}{=} {T^\alpha}_{\mu \nu}
{S_\alpha}^{\mu \nu}. \end{equation}

The action of $f(T )$ theory is defined as \begin{eqnarray} && {\cal L}({h^a}_\mu)=\int
d^4x |e|\left[\frac{1}{16\pi}f(T)+{\cal L}_{matter}\right], \end{eqnarray}
where $|e|=\sqrt{-g}=det\left({e^a}_\mu\right)$,
and $  {\cal L}_{ matter}$ is the Lagrangian of matter field. We
assume the units in which $G = c = 1$.   Considering the action in equation (7) as a
function of the fields ${e^a}_\mu$ and its first derivatives and putting  the variation of the
function with respect to the field ${e^a}_\mu$ and its first derivative  to be vanishing, one
can obtain the following equations of motion \cite{FF,FF1}: \begin{equation} \label{q8}
{S_\mu}^{\rho \nu} T_{,\rho} \
f(T)_{TT}+\left[h^{-1}{e^a}_\mu\partial_\rho\left(e{e_a}^\alpha
{S_\alpha}^{\rho \nu}\right)-{T^\alpha}_{\lambda \mu}{S_\alpha}^{\nu
\lambda}\right]f(T)_T+\frac{1}{4}\delta^\nu_\mu f(T)=4\pi {{{\cal
T}^{{}^{{}^{^{}{\!\!\!\!}}}}}}^\nu_\mu,\end{equation}
 where
$T_{,\rho}=\frac{\partial T}{\partial x^\rho}$, \ \
$f(T)_T=\frac{\partial f(T)}{\partial T}$, \ \
$f(T)_{TT}=\frac{\partial^2 f(T)}{\partial T^2}$ and ${{{\cal
T}^{{}^{{}^{^{}{\!\!\!\!}}}}}}^\nu_\mu$ is the
energy momentum tensor. Field equations (8) reduce to the TEGR field equation when  $f(T)=T$.

\newsection{$f(T)$ cosmological models and Phase-Space Diagram}
We apply the field equations (\ref{q8}) of  $f(T)=T+\tilde{\alpha} (-T)^n$ to the FLRW universe of a spatially homogeneous and isotropic spacetime \cite{R32}. Since $\tilde{\alpha}$ has length dimensions $2(n-1)$ we take $\tilde{\alpha}=\alpha^{2n-2}$. The tetrad field, for flat spatial curvature,  can be written in  polar coordinate ($t$, $r$, $\theta$, $\phi$) as follows:
\begin{eqnarray}\label{tetrad}
\nonumber \left({e_{i}}^{\mu}\right)=\left(
  \begin{array}{cccc}
    1 & 0 & 0 & 0 \\
    0&\frac{\sin{\theta} \cos{\phi}}{a(t)} & \frac{ \cos{\theta} \cos{\phi}}{r a(t)} & -\frac{ \sin{\phi}}{ r a(t)\sin{\theta}} \\[5pt]
    0&\frac{ \sin{\theta} \sin{\phi}}{ a(t)} & \frac{ \cos{\theta} \sin{\phi}}{r a(t)} & \frac{ \cos{\phi}}{r a(t)\sin{\theta}} \\[5pt]
    0&\frac{ \cos{\theta}}{a(t)} & \frac{- \sin{\theta}}{r a(t)} &0 \\[5pt]
  \end{array}
\right).&\\
\end{eqnarray}
 The equation of state (EoS) is assumed for an isotropic fluid such that the energy-momentum tensor  takes the form ${{\cal T}_{\mu}}^{\nu}=\textmd{diag}(\rho,-p,-p,-p)$. The tetrad (\ref{tetrad}) has the same metric as FRW metric. Applying equation (\ref{tetrad}) to the  $f(T)$ field equations (\ref{q8}) read
\begin{eqnarray}
  \rho   &=& \frac{1}{16\pi}(f+12H^2 f_T),\label{TFRW1} \\
  p &=&  \frac{-1}{16\pi}\left[(f+12H^2 f_T)+4\dot{H}(f_T-12H^2f_{TT})\right], \label{TFRW2}
\end{eqnarray}
where $\rho$ and $p$ are the total density and pressure of the matter inside the universe, $H$ is the Hubble parameter and $\dot{H}$ is the rate of Hubble parameter. Using Eq. (\ref{TFRW1}), Eq. (\ref{TFRW2}) can be rewritten as
\begin{equation}\label{t1}
p+\rho=-\frac{\dot{H}}{4\pi}(f_T-12H^2f_{TT}), \qquad \Rightarrow \dot{H}=-\frac{4\pi(p+\rho)}{(f_T-12H^2f_{TT})}.
\end{equation}
Using equation (\ref{tetrad}), the scalar torsion and the continuity equation take the form
\begin{equation}\label{t2}
T=-6H^2, \qquad \dot{\rho}=-3H(p+\rho).
\end{equation}
Now we are going to use the EoS in the form $p=(\bar{\omega}-1)\rho$. From (\ref{t2}) and (\ref{TFRW1}) in (\ref{t1}) we get
 \begin{equation}\label{t3}
\dot{H}=-\frac{\bar{\omega}(f+12H^2f_{T})}{4(f_T-12H^2f_{TT})}.
\end{equation}
Using the general form of $f(T)=T+\tilde{\alpha} (-T)^n$ in (\ref{t3}) we get
\begin{equation}\label{t7}
 \dot{H}=-\frac{3\bar{\omega} H^2[1-(2n-1)\,({\bar\alpha} H)^{2(n-1)}]}{2[1-n(2n-1)\,({\bar \alpha} H)^{2(n-1)}]}, \qquad {\bar{\alpha}}={\alpha}\sqrt{6}.\end{equation}
Now before presenting a general solution for this equation and its limitations let us briefly present possible branches of solutions using phase-space method. This is going to help us to understand when this solution form is valid. Let us define the following dimensionless quantities, $\tau$ and ${h}$
\begin{equation} { h}=[n (2n-1)]^{\frac{1}{2n-2}}\, \bar{\alpha} H, \quad  { \tau}={3\bar{\omega} \over 2} \, n^{\frac{2n-1}{2-2n}}\,[(2n-1)\, ]^{1 \over 2-2n}\,t/\bar{\alpha}.\end{equation}
 \begin{equation}\label{t7.5}
 { dh \over d\tau}=F(h)=-\frac{ { h}^2(n-{ h}^{2(n-1)})}{1-{ h}^{2(n-1)}}.
\end{equation}
\begin{figure}
 \centering
  {{\includegraphics[angle=-90,width=82mm]{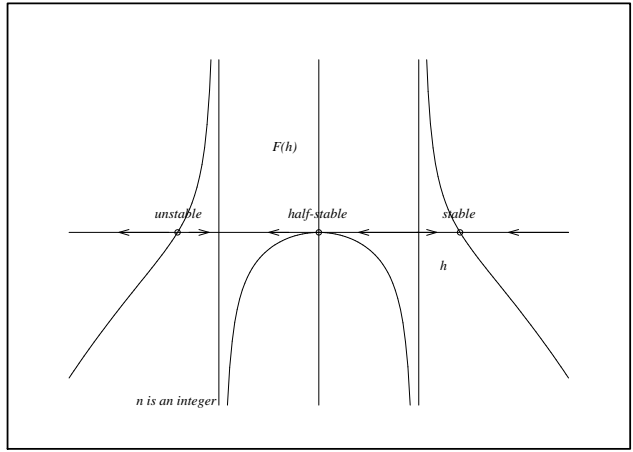}}}
 \caption{\footnotesize A typical phase diagram for integer $n$, where $n=2$, which shows types of fixed points and time evolution direction.}
 \label{fig-2}
 \end{figure}

We are going to follow the analysis introduced in \cite{awad1} to describe different branches of solutions for this case. An important input in this analysis is the phase-space diagram showing $dh/d\tau$ as a function of $h$, $F(h)$, and the positions of fixed points. Fixed points are values of h that satisfy $F(h) =0$, which we will call, $h_i$'s. These $h_i$'s are solutions of Eq. (\ref{t7.5}) which describe de Sitter cosmology. If the system started at one of these values,  $h_i$'s, it will stay at this value forever. But if it started from a value close to say $h_i$, it might move towards the fixed point or move away from it, which we classify as stable or unstable fixed points, respectively. Also, there is a notion of half-stable fixed point which works as a stable point from one direction and as an unstable from the other. We shown the fixed points for $n=2$ case as an example in figure (1).

Using these phase-space diagrams one can show how solutions evolves. For example, let us assume that the universe started at $h(0)=1/2$, from the flow directions in figure (2-a), $h(\tau)$ evolves to the left till it hits the fixed point at $h=0$ after infinite time. One of the important properties of these fixed points is that they are reached after an infinite time. This can be shown easily by integrating equation (\ref{t7.5}) using a new variable $\bar h= h^{2n-2}$, then we get
\be \tau= \lim_{\bar h\rightarrow { n }}\int_{\bar h}^{ h*} {\bar h}^{1-2n \over 2n-2}\left[{1-{\bar h} \over {n-{\bar h} }}\right] d{\bar h}= \lim_{\bar h\rightarrow {n }} \left[ C_1+C_2\log({\bar h}-n)+C_3 ({\bar h}-n)+ ... \right]. \ee This expression diverges in the limit because of the logarithmic term leading to an infinite time at the fixed point $\bar h=h^{2n-2}=n$.

In some cases, parts of the phase diagram could lie between two fixed points in this case these parts consume the whole time of evolution and constitute a branch of solution. In other cases, evolution could stop because of a curvature singularity, as in figure (2-a) at $h=1$. In general, knowing how different parts of a solution evolve helps us to divide it into different branches. Another important feature in figure (2-a) is the existence of a sudden singularity at $h=1$, since $dh/d\tau \rightarrow \infty$ as $h\rightarrow 1$ and one can show that $a\neq 0$ in this limit.

For the case where $n$ is an integer, it is natural to divide the phase diagram into different regions, three for $h>0$ and three for $h<0$ as in figure (2-a). Notice that time evolution can not take $h$ from any value in these branches to another in a different branch. In this case for $h>0$ we have three regions $i$ (for $0<{ h}\leq 1$), $ii$ (for $1 <{ h}\leq \sqrt{2}$) and $iii$ ($\sqrt{2}<{ h}$). In region $i$ the universe is expanding and evolving towards an empty universe which is matching the standard FLRW cosmological model (or toward a de Sitter if we add a cosmological constant). Furthermore, in this region, the universe starts from a sudden singularity, instead of a big bang singularity, at earlier time (the point where $ h=1$). As Eq. (\ref{t7.5}) shows, the value of ${\bar \alpha}$ is important for identifying which branch matches our universe. Let us explain this following figure (2-a). If we take ${\bar \alpha}\sim G^{n-1}$, then our universe matches region $i$ since ${H} < 1/{\bar \alpha} \sim l_p{}^{-1}$ and is evolving from a sudden singularity where ${H} \sim 1/{\bar \alpha} \sim l_p{}^{-1}$, which is very close to the Planck scale. Region $ii$ ($1<{\bar h}\leq \sqrt{2}$), describes an inflationary phase in which both acceleration and ${\dot h}$ are positive. Another possible choice which is adopted here although it is common among authors studying $f(T)$ cosmology (for example see \cite{CCLS}) is to choose the value of ${\bar \alpha}$ to be related to the cosmological constant, $\Lambda$, since at the fixed point ${H} \sim 1/\sqrt{2 \alpha} \sim \sqrt{\Lambda}$. In this case, region $iii$ ($\sqrt{2}<{ h}$), describes our universes today with the usual big bang singularity at very early times but we have no access to regions $i$ and $ii$, since it will take our universe infinite time to reach the fixed point at ${H} \sim 1/\sqrt{2 \alpha} \sim \sqrt{\Lambda}$.

Notice that in case of integer $n$ the phase diagram is symmetric as in figure (2-a), while for half-integer $n$ the diagram has no symmetry as in figure (2-b). The existence of region like $i'$ as in figure (2-a) is going to play an important role in extending the spacetime beyond the sudden singularity as we will see in the next sections. Notice also  that for real $n$ (n>1), as in figure (3), generally we don't have any branches of solutions for $h<0$ which will make these spacetimes difficult to extend unless we introduce another equation of state in this region.

\begin{figure}
 \centering
  {{\includegraphics[angle=-90,width=70mm]{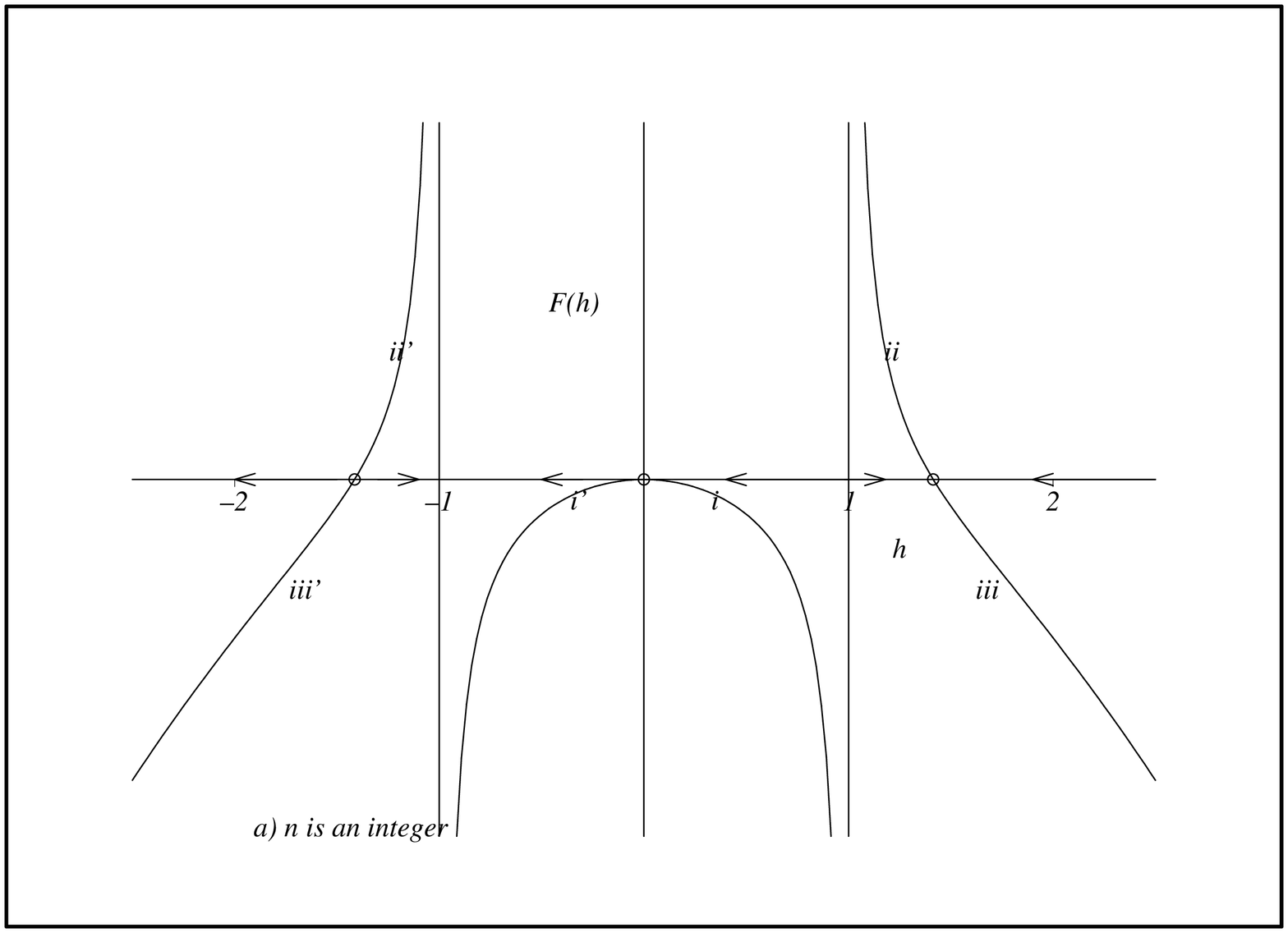}}{\includegraphics[angle=-90,width=70mm]{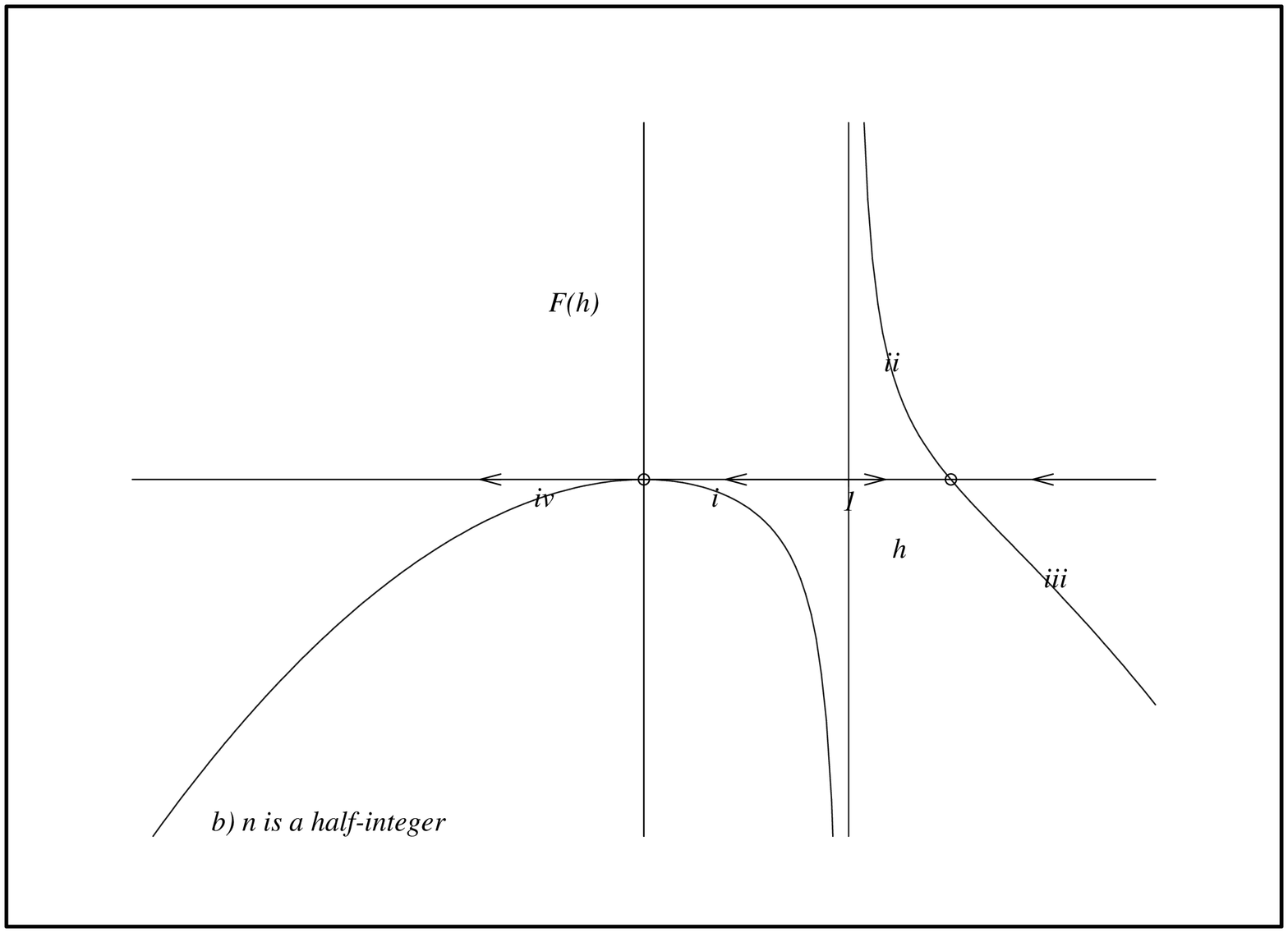}}}
 \caption{\footnotesize {\bf a)} The phase diagram for integer $n$, where $n=2$. It shows different branches of solutions. {\bf b)} The phase diagram for half-integer $n$, where $n=3/2$. It shows different branches of solutions.}
 \end{figure}
\newpage
 \begin{figure}
 \centering
  {{\includegraphics[angle=-90,width=70mm]{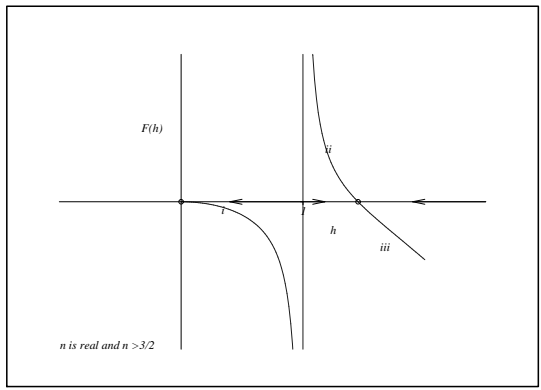}}}
 \caption{\footnotesize The phase diagram for real $n$, where $n>3/2$. It shows fixed points and time evolution directions.}
 \label{fig-4}
 \end{figure}

\subsection{A general solution and its limitations}
In this subsection we discuss the general solution presented above and its limitations.
Solving equation (\ref{t7}) for form $f(T)=T+\tilde{\alpha}(-T)^n$, one can obtain the following general solution in terms of hypergeometric function $_2F_1$
\begin{equation}\label{t8}
\tau({ H})=\frac{2}{3{\bar{\omega}}}\left[const.+H^{-1}\left[n-(n-1)\, _2F_1(1,{1 \over 2-2n};{2n-3 \over 2n-2};(2n-1)({\bar \alpha} H)^{2n-2})\right]\right].\end{equation}

First, using the known properties of the generalized hypergeometric functions, one can show the following;\\

i)\, For $n=1$, the solution is not defined since the parameter in the second list is infinite.\\

ii)\, For $n=1/2$, the solution is well defined since $_2F_1(1,1;2;0) \rightarrow 1$, therefore,
\be H(t)={1 \over 3/2 (\bar{\omega}+1)\,t+C'}.\ee\\

iii)\, For $n=3/2$, the solution is everywhere singular for $2{\bar \alpha}H < 1$, which is the region we are interested in this work as we will see below. Notice that cases with $n=0,1/2,1$ are nothing but TEGR cosmology with or without a cosmological constant. The first non trivial case is the one with $n=3/2$ which we are going to discuss it together with $n= 2$ which represent the general cases with integer or half-integer "$n$" in our discussion.

iv) As we take $\bar{\alpha} \rightarrow 0$ the solution tends to FLRW solution in teleparallel gravity (i.e., $f(T)=T$)
\begin{equation}\label{t9}
\lim_{\bar{\alpha} \rightarrow 0} \,_2F_1(1,{1 \over 2-2n};{2n-3 \over 2n-2};(2n-1)({\bar \alpha} H)^{2n-2})=1,\end{equation}
which leads
\begin{equation}\label{t10}
\tau({ H})=\frac{2}{3{\bar{\omega}}} \left[const.+H^{-1}\right],\end{equation}

v) If $p$ is the number of parameters in the first argument of the hypergeometric function (in our expression, $_2F_1(1,1/(2-2n);2n-3/(2n-2);z)$, this number is $2$) and q is the number of parameters in the second argument (this number is $1$). If no nonpositive integer exists in these two lists and $p=q+1$, the series converges only for $z<1$ and the series has a branch point at $z=1$ and a branch cut for $z \in (1,\infty)$ \footnote{One can check rigourously the divergence of this series at $z=1$ from the condition for convergence which states that $Re [ \,\sum_i \, b_i-\sum_k a_k] >0$ which in this case is vanishing, where $a_i$'s and $b_i$'s are the list parameters in $_pF_q(a_1,a_2,..;b_1,b_2,..;z)$. }. The above expression for $t(H)$ diverges not only for $ (2n-1)({\bar \alpha}H)^{(2n-2)}=1$ but also at $H=0$, this is consistent with the fact mentioned above that these two points are fixed points and only reached after infinite time.

\subsection{Specific solutions with different values of $n$}
Here we present explicit solutions for  $n=3/2, 2$ as representative cases for these cosmologies described here. We will see in the coming sections that cases with integer and half-integer $n$ are qualitatively different.\\
{\bf I- n = 3/2 case}\\
In this case we have
\begin{equation}\label{t15}
 \frac{d{ h}}{d{{ \tau}}}=-\frac{ { h}^2(3/2-{  h})}{1-{ h}},\qquad \qquad  \frac{\dot \eta( \tau)}{\eta( \tau)}=\gamma  h( \tau), \qquad \gamma^{-1}=\frac{1}{ \omega}.\end{equation}
 Equation (\ref{t15}) has the following solution
 \begin{eqnarray}\label{t177}
\A \A h( \tau)=\frac{1}{\frac{1}{3}\exp\left\{-W\left(-e^{\frac{-(9\tau+2)}{2}}\right)-\frac{9 \tau+2}{2}\right\}+c_1}\label{t16}, \nonu
 \A \A  \eta( \tau)=c_2 W\left(-e^{\frac{9 \tau+2}{2}}\right)^{-\gamma/9}\left[W\left(-e^{\frac{9 \tau+2}{2}}\right)-2\right]^{-\gamma/3}\label{t17}, \end{eqnarray}
where $W$ is the  Lambert $W$ function and the integration constant $c_1=\frac{2}{3}$ after using the initial condition $h(0)=1$  in Eq. (\ref{t16}). Also, equation (\ref{t16}) gives the value of $c_2=W\left(-e\right)^{\gamma/9}\left[W\left(-e\right)-2\right]^{\gamma/3}$  after using the initial condition $\eta(0)=1$.
It is important to notice that this solution does not have a time reflection symmetry.

{\bf II- n =2 case}\\
In this case we have
\begin{equation}\label{e2}
 \frac{d{h}}{d{{ \tau}}}=-\frac{ { h}^2(2-{  h}^2)}{1-{ h}^2},\qquad \qquad  \frac{\dot \eta( \tau)}{\eta( \tau)}=\gamma  h( \tau), \qquad \gamma^{-1}=\frac{4}{3\bar \omega}.\end{equation}
 Equation (\ref{e2}) has the following solution
 \begin{eqnarray}\label{t17}
t={1 \over 2} {1 \over { h}(  \tau)}+ {\sqrt{2} \over 4} \tanh^{-1}\left({{ h}(  \tau) \over \sqrt{2}}\right)+C_1. \end{eqnarray}
Using the continuity equation and equation (10) we obtain
\be { h} ( \tau)=\pm \sqrt{1\mp \sqrt{1-\eta^{-3 \bar \omega}}}.\ee From equation (25) we get
\be  \tau=C_2\, \pm \left[ {\sqrt{2} \over 4} \tanh^{-1}({1\over \sqrt{2}}\sqrt{1\mp \sqrt{1-\eta^{-3 \bar \omega}}})+{1 \over 2 \sqrt{1\mp \sqrt{1-\eta^{-3 \bar \omega}}}}\right].\ee
Notice that this solution has a time reflection symmetry which is going to provide us with natural extension for the spacetime beyond the singular point as we will see in the next section.
It is intriguing to notice that this solution is identical to the solution found in \cite{awad2} where quantum corrections due to Weyl anomaly generated higher-curvature terms which left the initial singularity traversable. This clearly shows that higher-curvature terms indeed affect singularities in gravitational theories.

 \section{Junction analysis}
 In this section we are going to find the general solution near the singularity in order to join two disjoint patches of solutions. We apply junction conditions for $f(T)$ theories derived in \cite{DDG} to show that the extended spacetime is consistent with the field equations of this higher-curvature teleparallel gravity. As we will see in the next section this naturally introduces extension to nonspacelike curves beyond the singularity through joining two branches of solutions.
 \subsection{General solution near singularity}
Here we present the form of a general solution in the case where "n" is an integer.
\begin{figure}
\begin{center}
{\includegraphics[angle=-90,width=82mm]{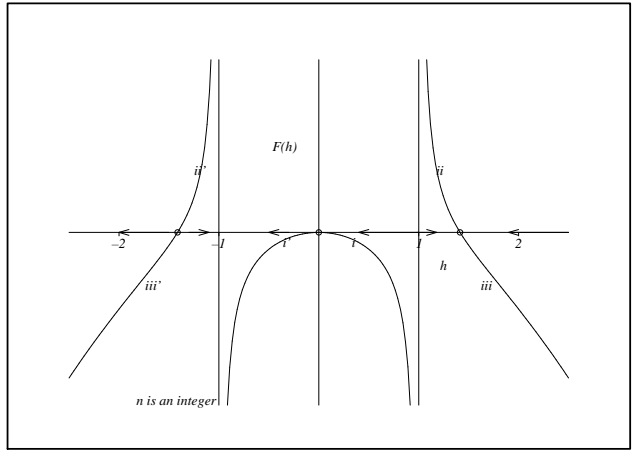}}
\caption{Different branches or regions of solution for integer n; $i$, $ii$, $iii$, $i'$, $ii'$, and $iii'$.}
\end{center}
\end{figure}
Solving Eq. (\ref{t7.5}) around $ h=1$ we get
\begin{equation}\label{t18}
 h( \tau)=1\mp\sqrt{1+ \tau+c_3}, \end{equation}
Equation (\ref{t18}) gives the value of $c_3=-1$ when $ h(0)=1$. Then Eq. (\ref{t18})  takes the form \begin{equation}\label{t19}
 h( \tau)=1-\sqrt{ \tau}, \qquad {\textrm for} \qquad  \tau>0 \Rightarrow  h>0.\end{equation}
Using the fact that
 \begin{equation}\label{t19}
\frac{\dot \eta}{\eta}= {\gamma}{ h ( \tau)}\Rightarrow \eta( \tau)=c_4[1+\gamma  \tau-\frac{2}{3} \gamma  \tau^{3/2}+O\left( \tau^2\right)], \end{equation}
which is a good description for region $i$ near $h=1$ and $\tau >0$. For integer $n$ Eq. (\ref{t7.5}) admits another solution using its symmetric time $ \tau \rightarrow -\tau$, $ h\rightarrow - h$
 \begin{equation}\label{t19}
 h( \tau)=-1+\sqrt{- \tau}, \qquad {\textrm for} \qquad  \tau<0 \Rightarrow  h<0.\end{equation}
Using the above expression for $h$, one can obtain an expression for the scale factor $\eta$ which reads
 \begin{equation}\label{t19}
 \eta( \tau)=[1-\gamma  \tau-\frac{2}{3} \gamma (-\tau)^{3/2}+O\left( \tau^2\right)], \end{equation}
which is a good description for region $"i' "$ near $h=-1$ and $\tau <0$.

Here it is constructive to comment on our local solution and our cosmological model in general which is based on choosing region "$i$" to describe our universe. In fact, this is a consequence of choosing $\tilde {\alpha} \sim G^{n-1} $. Most solutions in literature  (see for example \cite{FF1}) consider region $iii$ to describe the universe since their choice of  $\tilde {\alpha}$ is related to the value of the cosmological constant as it is clear from Figure (4). In this case, a local solution around the big bang singularity is governed by \be {\dot H}= -{3 {\bar w} \over 2n}H^2+ O(H^{2-2n}), \ee for $n>1$, which has the following solution \be a(\tau)=a_0\,\tau^{2n \over 3 {\bar w}}.\ee Clearly this is different from the expression in (\ref{t19}). We emphasis here that most of the model as in \cite{FF1} uses region "$iii$" to describe the universe while our model uses region "$i$" to describe the universe which could be useful in studying the initial singularity at $t=0$ as we will see below.

 Now we can join two branches of solutions together, namely, region $"i "$ and $"i'"$ at $\tau =0$ to form a spacetime defined for all values of $\tau$. In this case, the scale factor of this spacetime is given by
 \begin{equation}\label{t19-2}
 \eta( \tau)=[1+\gamma | \tau|-\frac{2}{3} \gamma | \tau |^{3/2}+O\left( \tau^2\right)]\end{equation}
 Such a solution represents our extended spacetime and will be considered in the following sections to show that {\it a)} this junction is consistent with the field equations, {\it b)} one can use this spacetime to extend timelike and lightlike curves beyond the singular point, therefore, showing the singularity is traversable.
 \subsection{Junction condition}
Here we apply the junction conditions of the higher derivative gravitational
theory given by equation (8) following the work in \cite{DDG}. Using Gaussian normal coordinates near a
hypersurface $\sum $ with a metric ${\tilde g}$, the line element takes the form
 \begin{equation}\label{t20}
 ds^2=-dw^2+{\tilde g}_{ij} dx^idx^j.\end{equation}
 Junction condition for the $f(T)$ theories have been analyzed in \cite{DDG} where
 \begin{equation}\label{t21}
 \left[\tilde g_{ij}\right]^+_-=0, \qquad   \left[{\tilde e}^a{}_i\right]^+_-=0, \qquad {\textrm and} \qquad  \left[f_T(\Theta_{ij}-\tilde g_{ij} \Theta)\right]^+_-= \left[ T_{ij} \right]^+_-=0,\end{equation}
where \begin{equation}\label{t21} \Theta_{ij}=-H a^2\delta_{ij} \Rightarrow \Theta_{ij}-\tilde g_{ij} \Theta=2Ha^2\delta_{ij},\end{equation}
and \begin{equation}\label{t22} \tilde g_{ij} \Theta^{ij}=-3H, \qquad f_T=1-{ n} (\bar \alpha H)^{2(n-1)}, \hspace{0.2 in}{\textrm where \; n\; is\; an\; integer} \end{equation}
\begin{eqnarray}\label{t22} \A \A \left[2\Biggl\{1-{ n} (\bar \alpha H)^{2(n-1)}\Biggr\}Ha^2\delta_{ij}\right]^+_-= \left[ T_{ij} \right]^+_-=0,\nonu
 \A \A \Rightarrow \left[ T^j_i \right]^+_-={8(n-1)\,n^{\frac{1}{2(1-n)}}{\bar \alpha}\,}\delta^j_i, \quad {\textrm where} \quad H_0= n^{1 \over 2(1-n)} {\bar \alpha}.\end{eqnarray}
Equation (\ref{t22}) indicates that pressure is going to be \begin{equation}\label{t23} p \rightarrow p+p_0 \delta(t), \qquad p_0={8(n-1){ n}^{\frac{1}{2(1-n)}} {\bar \alpha}}.\end{equation} The result shows that there is a delta Dirac function in the pressure which is needed to account for the jump in the extrinsic curvature component $\Theta_{ij}$. This result is very similar to the calculation of the junction condition in the case of Weyl anomaly in GR which produces a higher-curvature terms as a result of quantum corrections \cite{awad2}.

\section{Equation of motion of a test particle}
In teleparallel gravity equation of motion of a test particle replaces geodesic equations in GR. Gravity in teleparallel theories is no longer described by curvature of the spacetime but through torsion which can be thought as a force field defined all over spacetime. It is well known that this equation of motion is completely equivalent to geodesic equation in GR, this is why teleparallel gravity is equivalent to GR not only from field equation point of view but also from equation of motion of a test particle under the influence of gravity.
Since geodesics equations in general relativity are identical to equations of motion of test particles in teleparallel gravity it is natural to define a spacetime singularity in teleparallel gravity as points where timelike or lightlike curves end. Therefore, if we are able to extend timelike or lightlike curves in teleparallel theories beyond the singular point, this singularity can be called traversable. In the following analysis we show that it is possible to extend timelike or lightlike curves beyond the singular point which makes the singularity under consideration traversable.

These equation are:
 \begin{eqnarray}\label{t23}
\A \A  \frac{d^2x^i}{d\lambda^2}=2H\frac{dt}{d\lambda}\frac{dx^i}{d\lambda},\nonumber\\
\A \A  \frac{d^2t}{d\lambda^2}=a^2H\left(\frac{dx^i}{d\lambda}\right)^2. \end{eqnarray}
Equations (\ref{t23}) have the following solution
\begin{eqnarray}\label{t24}
\A \A  \frac{dx^i}{d\lambda}=\frac{p^i}{a^2}=f^i(\lambda),\nonumber\\
\A \A  \frac{dt}{d\lambda}=\pm\sqrt{\epsilon+\frac{p^2}{a^2}}=g(t). \end{eqnarray}
where $\epsilon$ and $p^i$ are constants of integration and $\lambda$ is a parametrization for the curve produced by the motion of the test particle. The tangent of the curve $u^a$ shows if the curve is timelike or spacelike and $\epsilon=-u^au_a$. $\epsilon$ is 1 if the tangent is timelke and 0 if it is  lightlike.

It is important to notice that Picard-Lendel$\ddot{o}$f theorem shows that there is a unique solution to Eq. (\ref{t24}) iff $g$ and $f^i$ are continues in $\lambda$ and Lipshitz continues in $t$. This is in fact grantee that timelike and lightlike curves can be extended across the singular point at $t=0$ as long as $u(0)$ does not  vanish and $H(0)$ has a finite discontinuity. This is the main mathematical result we rely on for extending timelike and lightlike curves across $t=0$.
\begin{equation}\label{t25}
a(t)=a_0[1+H_0\mid t \mid-\frac{2}{3} \gamma^{-1}\mid H_0{}t \mid^{3/2}+O(t^2)],\end{equation}  where $a(0)=a_0$. From the second equation of (\ref{t24}) we get
\begin{eqnarray}\label{t26}
\lambda(t)= \frac{t}{\chi}+sign(t)\,\frac{H_0 p^2a_0{}^2\,t^2}{2\chi^3}+O(t^3),\end{eqnarray}  where we have used $\lambda(0)=0$, and $ \chi=\sqrt{\epsilon +\frac{p^2}{a_0{}^2}}$.  Equation (\ref{t26}) gives
\begin{equation}\label{t27}
t(\lambda)=\chi \, \lambda-sign(\lambda) \frac{H_0^2 p^2\lambda^2}{2a_0{}^2}+O({\lambda}^3).\end{equation}
Integrating the first Eq. of (\ref{t24}) we get
\begin{equation}\label{t27}
x^i(\lambda)=x^i{}_0+\frac{p^i}{a_0{}^2}\,\lambda-sign(\lambda)\frac{H_0p^2\chi}{a_0^2}  \lambda^2+O({\lambda}^3),\end{equation} where $x^i(0)=x^i{}_0$.
Few comments here in order; first, these curves are $C^1$, therefore, the discontinuity in $\ddot a$ does not affect them. Second, the leading behavior in the parameter $\lambda$ is independent of the equation of state since it does not depend on $\gamma$ which insures that the leading behavior of these physical curves around the singularity is independent of the form of the equation of state.
\section{Conclusion}
Higher curvature terms in gravity can affect the behavior of gravitational solutions near regions with large curvatures since they spoil local energy conditions required for gravitational focusing that causes singularities \cite{frolov,frolov1}. It is interesting to show examples of gravitational theories with high-curvature/torsion terms that can affect spacetime singularities and leave them traversable. One of the interesting theories that can contain such higher-curvature/torsion terms and yet equivalent to GR in the small curvature/torsion limit is the generalized teleparallel gravity or $f(T)$ gravity. In this work we present a class of new cosmological solutions for $f(T)$ theory with $f(T)=T+\tilde{\alpha} \,(- T)^n$, using the usual equation of state $p=(\bar{\omega}-1)\, \rho$, where $\tilde{\alpha}$ is some parameter and $n$ is an integer. These solutions have an exact form and approaches the standard cosmological model at late times. We show that for a certain class of models (where n is an integer), higher-curvature/torsion corrections in teleparallel theories change the nature of the singularity from a big bang to a sudden singularity. Geodesics of spacetimes in GR can be extended beyond sudden singularities, here we also show that timelike or lightlike curves in teleparallel theories can be extended across the singular point leaving these sudden singularities traversable. In order to check the consistency of these extensions with the field equations it is essential to show that the extended spacetime satisfies the known junction conditions of $f(T)$ theories which has been derived in \cite{DDG}. This suggests that the expansion phase has been preceded by a contraction phase. It is interesting to notice that this solution is identical to the solution found in \cite{awad2} where quantum corrections due to Weyl anomaly generated higher-curvature terms which render the initial singularity traversable. This clearly shows that higher-curvature/torsion terms affect singularities in gravitational theories.

\end{document}